\documentclass[sigconf]{acmart}
\AtBeginDocument{%
  }

\setcopyright{acmlicensed}
\copyrightyear{2018}
\acmYear{2018}
\acmDOI{XXXXXXX.XXXXXXX}
\acmConference[Conference acronym 'XX]{Make sure to enter the correct
  conference title from your rights confirmation email}{June 03--05,
  2018}{Woodstock, NY}
\acmISBN{978-1-4503-XXXX-X/2018/06}

\usepackage[most]{tcolorbox} % Added to include Feedback C  
\newtcolorbox[auto counter]{llmprompt}[2][]{float,  boxsep=1pt,before skip=0pt,
  after skip=0pt,title=Prompt~\thetcbcounter: #2,#1}
\newtcolorbox[auto counter]{llmresponse}[2][]{float, boxsep=1pt,before skip=0pt,
  after skip=0pt,title=Response~\thetcbcounter: #2,#1}
\newtcolorbox[auto counter,number format=\Alph]{feedback}[2][]{title=Feedback~\thetcbcounter: #2,#1,breakable} % removed float, added breakable

\newcommand{\promptinsert}[1]{\texttt{{\textless}#1{\textgreater}}}
\usepackage{xcolor} % added by S
\usepackage{xcolor}
\usepackage{tikz}
\usepackage{subcaption}  % for side-by-side subfigures
\usetikzlibrary{shapes.geometric, arrows,positioning}
\usepackage{pgfplots} 
\usepgfplotslibrary{statistics}
\pgfplotsset{compat=1.18}
\usepackage{caption}
 \definecolor{lightblue}{RGB}{180, 180, 220}  
\definecolor{pastel}{RGB}{151, 188, 203} 
\definecolor{darkerblue}{RGB}{0, 100, 130} 

\usepackage{tabularx}
\usepackage{caption}
\usepackage{xtab}
\usepackage{listings}
\usepackage{array}
\usepackage{longtable}
\usepackage{enumitem} % space before list
\setlist[itemize]{topsep=0pt, noitemsep, partopsep=0pt}
\usepackage{caption}
\begin{document}

\title{Fine-Tuning Models for Automated Code Review Feedback}

\author{Smitha Kumar}
\authornote{all authors contributed equally to this research.}
\email{smitha.kumar@hw.ac.uk}
\orcid{0000-0002-7583-1795}
\authornotemark[1]
\email{smitha.kumar@hw.ac.uk}
\affiliation{%
  \institution{Heriot-Watt University}
  \city{Dubai}
   \country{United Arab Emirates}
}
\author{Michael Adam Lones}
\orcid{0000-0002-2745-9896}
\affiliation{%
  \institution{Heriot-Watt University}
  \city{Edinburgh}
  \country{United Kingdom}}
\email{m.lones@hw.ac.uk}

\author{Manuel Maarek}
\orcid{0000-0001-6233-6341}
\affiliation{%
  \institution{Heriot-Watt University}
  \city{Edinburgh}
  \country{United Kingdom}}
\email{m.maarek@hw.ac.uk}

 \author{Hind Zantout}
\orcid{0000-0002-3804-0513}
\email{h.zantout@hw.ac.uk}
\affiliation{%
  \institution{Heriot-Watt University}
  \city{Dubai}
   \country{United Arab Emirates}
}
\renewcommand{\shortauthors}{Trovato et al.}

\begin{abstract}
 Large Language Models (LLMs) have introduced new possibilities for programming education through personalized support, content creation, and automated feedback. While recent studies have demonstrated the potential for feedback generation, many techniques rely on proprietary models, raising concerns about cost, computational demands, and the ethical implications of sharing student code. Open LLMs provide an alternative approach, but they do not currently have the capabilities of proprietary models. To address this problem, we investigate whether parameter-efficient fine-tuning (PEFT) and prompt engineering---both of which distil knowledge from a dataset derived from a large, more capable model can be used to adapt and enhance the quality of feedback generated by the open LLM Code Llama. Feedback quality on buggy Java code was assessed using a combination of student evaluation, manual annotation and the automated metrics BLEU, ROUGE, and BERTScore. Our findings indicate that PEFT leads to notable improvements in feedback quality, and significantly outperforms prompt engineering, providing an avenue for developing freely-deployable feedback tools that can be effectively used to guide student learning. Student evaluation indicates that learners value the PEFT model’s feedback, and see it as being equally effective as the proprietary ChatGPT model. 
 %Unlike the proprietary model, it does not provide the solution directly, thereby promoting independent problem-solving. 
 Participants suggested that incorporating additional explanation for technical terms in the PEFT model's feedback could be more beneficial. This study demonstrates that fine-tuned models can effectively support critical thinking and guide the design of scalable pedagogical systems.
\end{abstract}



 % \includegraphics[width=\textwidth]{sampleteaser}
  %\caption{Seattle Mariners at Spring Training, 2010.}
  %\Description{Enjoying the baseball game from the third-base
 % seats. Ichiro Suzuki preparing to bat.}
  %\label{fig:teaser}


\maketitle

\section{Introduction}
Generative AI tools have significantly reshaped programming education in recent years. Prior to the emergence of Large Language Models (LLMs), automated feedback generation techniques were constrained by several technical challenges that hindered their broader adoption in educational settings \cite{10.1145/3764593}. Recent studies indicate that LLMs are capable of outperforming traditional feedback generation approaches~\cite{xia2023automated}, which has led to a significant increase in leveraging them for more precise, personalized, and effective feedback. According to~\citet{narciss2008feedback}, “feedback refers to all post-response information which  informs the learner on his/her actual state of learning or performance in order to regulate the further process of learning”. Feedback types include  
knowledge of performance (KP), knowledge of result or response (KR), and knowledge of the correct response (KCR). More elaborate forms of feedback,  such as knowledge about task constraints (KTC), knowledge about concepts (KC), knowledge about mistakes (KM), knowledge about how to proceed (KH), and knowledge about meta-cognition (KMC) offer deeper guidance to learners.
% Immediate and timely feedback  fosters a  positive learning environment~\cite{bluefix}. 

Students often rely on proprietary LLMs to receive prompt responses during programming education~\cite{su16031245}. However, such models are often costly~\cite{koutcheme2024opensourcelanguagemodels}, lack transparency, and offer limited customization options for specific downstream tasks~\cite{machado2025publicsecuregenerativeai}. Despite their benefits, LLMs sometimes generate hallucinations, which can confuse learners~\cite{Prather_2023}. 
Moreover, they may provide complete correct solutions without guiding learners through the problem-solving process. This can hinder the development of  critical thinking and problem-solving skills, which are some of the key competencies in programming education~\cite{10636140}. 
 
Proprietary LLMs are  black-box models that do not release the model weights, whereas open LLMs provides access to their model weights and can be freely deployed on local computer systems~\cite{xu2025positionopenclosedlarge}. Open LLMs allow fine-tuning to suit diverse needs, while also providing greater transparency, privacy, and control~\cite{machado2025publicsecuregenerativeai}. Therefore, harnessing the potential of open LLMs to suit pedagogical goals is an important step towards their widespread acceptance in educational settings.
Recent studies on automated feedback generation techniques report that the majority of tools generate KM type feedback, with limited support for KH type feedback, which offers more effective guidance to learners~\cite{,10.1145/3636515}.  
%10.1145/3231711,10.1145/3764593

This study aims to develop a novel framework based on PEFT to create high-quality actionable feedback for Java programming assignments.
%To the best of our knowledge, no prior work has proposed a similar solution.
We also investigate how this approach compares against prompt engineering, an approach that can be used to generate KM-type and KH-type responses without additional training. A key challenge in programming education research is the lack of publicly available datasets. Most existing studies do not release their datasets, which limits the validation of
their findings. To overcome this limitation, researchers have begun leveraging LLMs to generate diverse
datasets~\cite{leinonen2024llmitationsincerestformdata}. As LLMs are largely trained on datasets written by humans, this approach offers a viable alternative for dataset generation. For the purposes of this research, we created and shared publicly a new dataset containing code, KM-feedback, and KH-feedback.
This research study is guided by the following research question: 
   \textit{  How can large language models be optimized to generate high-quality targeted and actionable feedback?}
 
Our contributions include: 
\begin{itemize}
      \item  A novel PEFT framework to generate pedagogically structured feedback.
     \item Empirical evidence demonstrating that PEFT leads to better feedback than baseline prompt engineering strategies and performs comparably to proprietary models. Student evaluations indicate a preference for this feedback over ChatGPT  in a formative lab setting.
     \item A new dataset\footnote{\url{https://anonymous.4open.science/r/JCODE_KM_KH-4BEC}} of annotated Java programs with KM-KH responses covering diverse error types in Java programming.
     
  \end{itemize} 

This article is organized as follows. In Section~\ref{sec:background}, we discuss the use of LLMs in programming education and review key optimization strategies. Section~\ref{sec:methodology} outlines the research methodology. Section~\ref{sec:results-discussion} presents the findings of this study including its limitations.
%, and Section~\ref{sec:illustrative-examples} shows illustrative examples.
Finally, conclusions and recommendations for future work are covered in Section~\ref{sec:conclusions}.

\section {Background}
\label{sec:background}
\subsection{LLMs in Programming Education}
The programming education landscape has been profoundly impacted by LLMs that generate code. Existing literature reveals that LLMs are leveraged in programming education for tasks such as code generation, debugging, clarifying complex concepts, and providing personalized feedback. Existing research also shows that students generally hold positive attitudes towards the use of LLMs~\cite{10.1145/3633053.3633057,Ahmed2024Potentiality}.
LLMs help students learn independently at their pace, receiving instant feedback and explanations for coding challenges. This real-time assistance across multiple programming languages enhances the learning experience \cite{chen2023gptutorchatgptpoweredprogrammingtool}. 
%Multiple studies highlight that LLMs can effectively reduce instructor workload and serve as a virtual tool to support student learning~\cite{10.1145/3626252.3630880}. Although the student sentiments towards LLMs are positive, instructors worldwide remain concerned about their impact on computing pedagogy. This is largely attributed to the strong ability of  LLMs to generate accurate code for programming assignments~\cite{10.1145/3764593}, coupled with  student's over-reliance on these tools~\cite{10.1145/3626252.3630880}.  

 The proprietary models used in most of these studies are associated with financial constraints, making open LLMs a viable alternative~\cite{10.1145/3564721.3565955}. However, these models have yet to match the performance of proprietary models~\cite{10.1145/3657604.3664665}. While previous studies have demonstrated the potential and challenges associated with LLMs, their effectiveness depends greatly on how they are optimized for specific use cases. Prompt engineering and fine-tuning have emerged as some of the most promising techniques to enhance the performance of the  models on downstream tasks~\cite{PORNPRASIT2024107523}. The following section provides an overview of prompt engineering and parameter-efficient fine-tuning (PEFT).
 %The effectiveness of Llama 3 model as feedback generation tools for introductory Python programming assignments was evaluated in \cite{10.1007/978-3-031-98417-4_30}. The findings indicate that temperature and top-p parameters did not significantly impact the model's performance.  The suitability of prompt techniques is task-dependent: one-shot performs best for easy problems, zero-shot for complex tasks, and few-shot is least effective. As LLM feedback includes incorrect responses,  a supervised approach, in which instructors review and select comments  is recommended to ensure its usefulness.
\begin{comment}
Research studies indicate that overreliance  negatively affects the acquisition of essential skills like problem-solving and critical thinking, which are central to programming education~\cite{Qureshi_2023}.The key considerations when adapting LLMs for programming education are academic integrity, student over-reliance, trustworthiness of the results, suitability for novice learners, and ethical challenges. Researchers emphasize the need to prioritize algorithmic problem-solving over syntactic elements in programming education~\cite{10.1145/3626252.3630958}. \end{comment}
\\

\begin{comment}
The authors of  \cite{10.1145/3641554.3701791} evaluated the performance of multiple open-source and proprietary large language models in generating feedback for Python programming tasks and in assessing the quality of that feedback. The findings suggest that open-source LLMs closely match proprietary models in their ability to generate and evaluate programming feedback. While GPT-4o and similar proprietary models perform better, the results suggest that smaller models may be able to achieve significant improvements through proper fine-tuning. A study similar to ours \cite{solano2025narrowinggapsupervisedfinetuning} examined whether supervised fine-tuning  can improve the performance of small open-source (Qwen3-4B, Llama-3.1-8B, and Qwen3-32B) language models  in generating feedback for C programming tasks. The findings indicate that smaller models showed substantial improvements offering a potential pathway for educators to design specialized pedagogical tools.  \end{comment}
% Prompt engineering and fine-tuning are the most widely used techniques to improve the performance of the  models on downstream tasks.
 \subsection{Prompt Engineering}   
Prompt engineering involves guiding a pre-trained LLM to generate a desired output using natural language queries. Zero-shot prompting and in-context learning are the most commonly used prompting strategies. In zero-shot prompting, the model is provided with a direct natural language instruction to generate a response for a task, and relies only on its trained knowledge. In-context learning extends this approach by including additional examples along with the query. Importantly,these approaches do not update the model's internal parameters~\cite{PORNPRASIT2024107523}. %A well-structured prompt improves accuracy without additional model training. Prompt engineering-based studies on feedback generation have predominantly relied on proprietary GPT models. A recent study investigated the  capabilities of the GPT4 model in  generating feedback for Java programming assignments~\cite{10.1145/3649217.3653594}. Their findings indicate that the LLM consistently delivered accurate responses, and that feedback could be optimized to include style recommendations. Nevertheless, the presence of misleading responses highlighted the need for human supervision within a pedagogical setting.
\subsection{Parameter-Efficient Fine-Tuning} 
Fine-tuning has emerged as a practical approach for adapting large pre-trained code LLMs to various downstream tasks. It can be broadly classified into two types depending on the number of parameters updated: full fine-tuning and parameter-efficient fine-tuning (PEFT). Full fine-tuning potentially updates all model parameters and can require large amounts of training data and computational resources. In contrast, PEFT can be done using much more limited computational resources as it updates only a subset of the model parameters. This also helps to minimize the risk
of catastrophic forgetting. PEFT can be implemented using various approaches, including Low-Rank Adaptation (LoRA)~\cite{hu2021loralowrankadaptationlarge} and  Quantized LoRA (QLoRA)~\cite{dettmers2023qloraefficientfinetuningquantized}. %The Lora configuration parameters significantly influence the performance of the model. The parameter $r$ is the rank of the low-rank matrices and it controls the number of trainable parameters. A higher value of $r$ indicates more trainable parameters. The \textit{scaling factor}, lora\_alpha controls the impact of the adaptation on the model weights. According to~\cite{hu2021loralowrankadaptationlarge}, $2 * r$, where $r$ is the rank of the low-rank matrices is often considered effective. low-rank matrices to the transformer layers while keeping the base model weights frozen.  This approach reduces the number of trainable parameters and computational cost. 
LoRA adds trainable low-rank matrices to the LLM's transformer layers while keeping the base model weights frozen. This approach significantly reduces the number of trainable parameters and computational overhead without compromising the model's accuracy.
%Instead of updating the full set of model parameters, LoRA introduces trainable low-rank matrices into the model architecture. For a given weight matrix, $W_0 \in \mathbb{R}^{d \times k}$, where $d$ and $k$ are the output and input dimensions respectively, LoRA introduces an update matrix $\Delta W$ that is low-rank defined as:\[\Delta W = BA,\] where $B \in \mathbb{R}^{d \times r}$, $A \in \mathbb{R}^{r \times k}$, and $r \ll \min(d, k) $, where \textit{r} represents the rank of the low-rank matrices.
%LoRA has a number of configuration parameters which significantly influence the performance of the model. The parameter $r$ is the rank of the low-rank matrices and it controls the number of trainable parameters; higher value results in more trainable parameters. The scaling factor, \textit{lora\_alpha} controls the impact of the adaptation on the model weights. According to~\cite{hu2021loralowrankadaptationlarge}, setting this to approximately $2 * r$ is often considered effective. Furthermore, \textit{lora\_dropout} controls the application of dropout, a regularization technique used to reduce overfitting. 
Quantized LoRA (QLoRA), which we use in this research, builds on LoRA by incorporating quantization that further reduces the memory footprint without degrading  performance. It uses a 4-bit floating-point value to store the model weights, applies double quantization, and uses paged optimizers to dynamically manage memory during training to achieve optimal performance.  
\section {Methodology}  
\label{sec:methodology}
\begin{figure}
    \centering
    \includegraphics[width=0.48\textwidth]{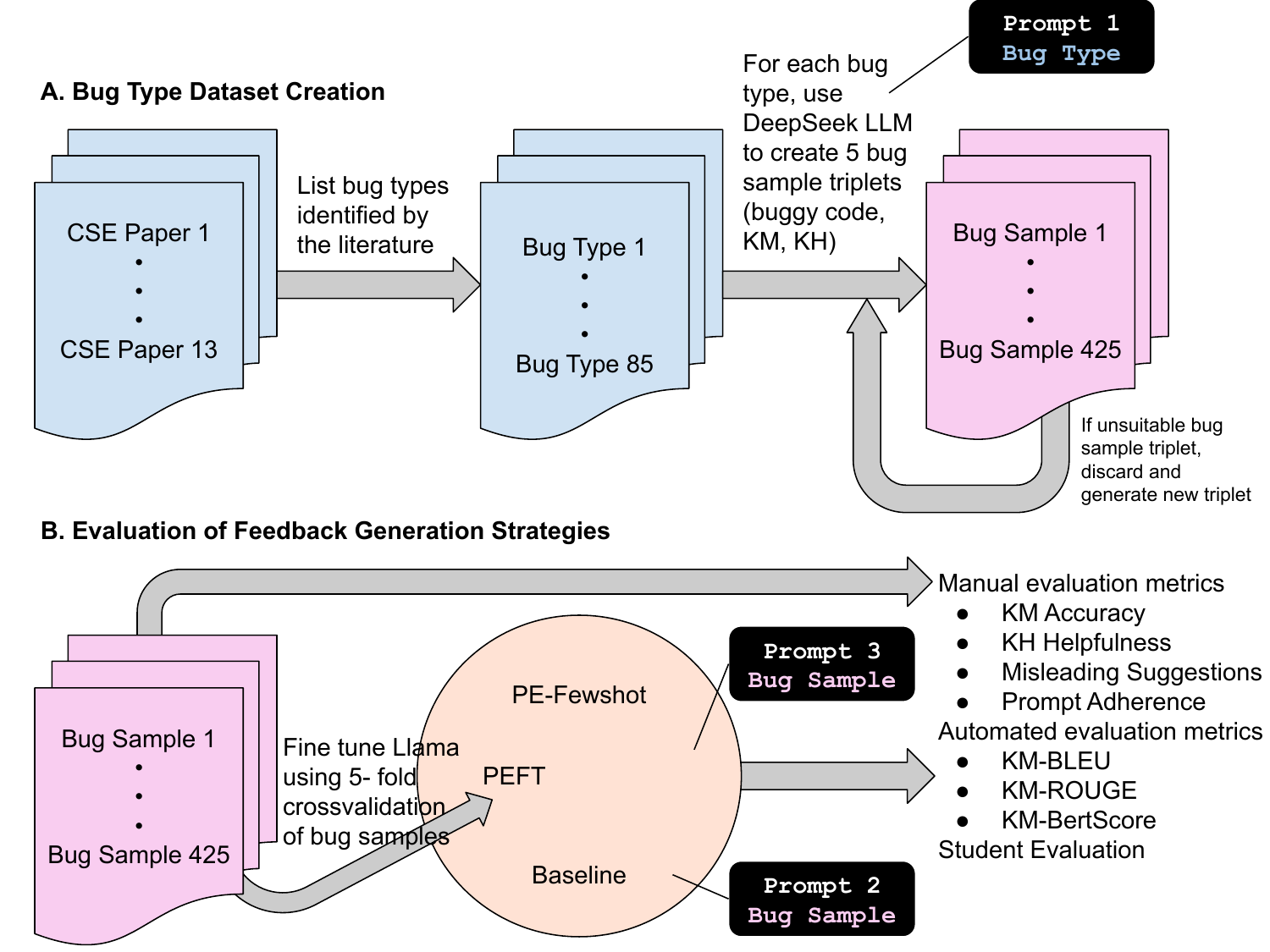}
   \Description{Flowchart showing the steps in A Bug Type dataset creation and B Evaluation of feedback generation strategies.}
    \caption{Overview of methodology, showing (A) the bug type dataset creation presented in Section~\ref{sec:dataset-creation}, and (B) the different feedback strategies presented in Sections~\ref{sec:PEFT} and~\ref{sec:prompt-engineering} and the evaluation metrics presented in Section~\ref{sec:evaluation-metrics}}
    \label{fig:methodology}
\end{figure}

In this section we explain the dataset creation process, model architecture and configurations, and the evaluation metrics. An overview of the methodology is shown in Figure~\ref{fig:methodology}.

\subsection{Dataset Creation}
\label{sec:dataset-creation}
Due to the scarcity of openly shared data in the programming education domain, researchers have investigated the ability of LLMs to generate synthetic data and found that the generated samples are not significantly different from actual student submissions~\cite{leinonen2024llmitationsincerestformdata}. To ensure that the synthetic dataset accurately reflects student errors, we analysed  the following research papers published in prominent computer science venues between 2005 to 2024, aiming  to identify common bug types frequently observed in student submissions~\cite{ 
10.1145/2676723.2677258,
10.1145/1067445.1067472,
7044420,
9462963,
10.1145/3335814,
10.1145/2632320.2632343,
10.1145/2325296.2325318,
10.1145/611892.611956,
10.1145/3077618,
10.1145/3160489.3160493,
1611967,
just2025bugsbreakthroughsnoviceerrors,
10.5555/3722479.3722540}.
We observed that certain errors were repeated across multiple papers. In some instances, errors with unclear or incomplete details were excluded. A total of 85 distinct bug types were identified, each type assigned a unique identifier (BugID) and recorded only once, regardless of its occurrence across multiple publications. The full list of bug types is available at \footnote{\url{https://anonymous.4open.science/r/JCODE_KM_KH-4BEC}}. The bug types served as the basis for generating the synthetic data in this study.
These include syntax, semantic, logical errors and conceptual misunderstandings commonly found in student submissions, and cover both compilation-time errors and runtime errors. To provide deeper insights into the model's capabilities, the bug types are classified into two categories: imperative (I), and object-oriented (O).
We employed prompt engineering with Deepseek-R1\footnote{\url{https://huggingface.co/deepseek-ai/DeepSeek-R1}} to generate labelled triplets in the form of \promptinsert{Code, KM-feedback, KH-feedback} to train our model. Deepseek-R1 was chosen for this study because its license permits  the model's output to be used in training other models. %The experiment was conducted using the following configuration: $\text{temperature} = 0.7$, $\text{maxtokens} = 81920$ and $\text{top-p} = 0.7$. 
Each bug type is illustrated with a corresponding code snippet that demonstrates a distinct code scenario. In each iteration, the placeholder \promptinsert{BUG\_TYPE} in Prompt~\ref{prompt:template-triplets} was replaced with a specific bug type (BugID) from \footnote{\url{https://anonymous.4open.science/r/JCODE_KM_KH-4BEC}}. For each bug type, the model generated five distinct triplets, which were subsequently validated by the first author in order to verify the correctness of the feedback. A total of 425 triplets were created for this study, containing 5 samples for each BugID in order to reflect natural variation in the occurrence of each bug. The generated feedback  serves as the ground 
 truth for this research.  %The dataset has been made publicly available\footnote{\url{https://anonymous.4open.science/r/JCODE_KM_KH-4BEC}}.
%\footnote{\url{https://github.com/ssk705/CodeLLM\_KM\_KH}}. 
This dataset has potential to support the development of pedagogical models,  and facilitate future studies.  A sample code snippet and the associated KM and KH response is shown in Response~\ref{response:triplet-example}.

 \begin{llmprompt}[label={prompt:template-triplets}]{{Prompt template for triplets (Code, KM, KH)}}
 \small
SYSTEM: You are an expert programming tutor with strong knowledge of common student mistakes from programming education literature. \\ 
USER: Your task is to generate five buggy Java code snippets that illustrate the common mistakes made by students in introductory undergraduate programming courses (CS1 and CS2 levels) aligned with the bug type specified below. Bug type: \promptinsert{BUG\_TYPE} \\ For each code provide: \\
** \textbf{KM (Knowledge about mistakes)} ** : Provide feedback that helps students understand their mistakes. If there are multiple mistakes, list them individually. \\
**\textbf{ KH (Knowledge about how to proceed)} ** : Provide guidance to the students on how to fix the mistakes. Do not provide the corrected code. Instead, please provide short, concise, and easily understandable explanations. \\ 
Here are a few examples:  \\
\promptinsert{JAVA CODE} \promptinsert{KM: {KM response}  KH: {KH response}}
\end{llmprompt}

\begin{llmresponse}[label={response:triplet-example}]{{Sample bug triplet}}
{\texttt{import java.util.*;\\
class Vehicle \{\}\\
class Car extends Vehicle \{\\
\mbox{\hspace{2em}}void startEngine() \{ \\
\mbox{\hspace{2em}}System.out.println("Vroom!"); \}
\} \\
public class Main \{ \\
\mbox{\hspace{2em}}public static void main(String[] args) \{\\
\mbox{\hspace{4em}}List<Vehicle> vehicles =\\
\mbox{\hspace{6em}}new ArrayList<>();\\
\mbox{\hspace{4em}}vehicles.add(new Car());\\
\mbox{\hspace{4em}}for (Vehicle v : vehicles) \{\\
\mbox{\hspace{6em}}v.startEngine(); \\
\mbox{\hspace{4em}}\}
\mbox{\hspace{2em}}\}
\}  }}\\
\textbf{Knowledge about mistakes (KM)} : Compile-time type of `v` is `Vehicle`, which lacks the `startEngine()` method.  Polymorphism doesn’t automatically expose subclass-specific methods without explicit casting.  \\
\textbf{Knowledge about how to proceed (KH)} : Cast `v` to `Car` inside the loop before calling `startEngine()`, ensuring `v` is a `Car` instance.
\end{llmresponse}

\subsection{PEFT-Driven Framework for Pedagogical Code Feedback}
\label{sec:PEFT}
 
 %Although proprietary models offer better performances due to extensive training on large datasets, they do not offer fine-tuning options. Fine-tuning is essential for adopting the model to task-specific requirements in a pedagogical setting.
  %The objective ofthis study is to optimize the model response to support learners; therefore, we adopted the open-source LLM, Code Llama. 
   Our framework builds on Code Llama-7b-Instruct-hf, a LLaMA causal language model based on the transformer architecture with 32 layers and support for very long context lengths. The framework's technical components include the base model, PEFT adapter module, input encoding and the training dataset. 
  Code Llama-7B-Instruct is based on the foundation model Llama 2~\cite{touvron2023llama2openfoundation} and is trained on 500B tokens and designed to follow human instructions. Additionally, it extends the context length from 4,096 tokens used in Llama 2 to 16,384 tokens. A longer context window enables the model to understand longer code samples, effectively process detailed prompts, and avoid truncations~\cite{rozière2024codellamaopenfoundation}.  While larger models offer significant performance benefits, the 7B model is favored for its efficient resource usage, ease of deployment in typical educational environments, and faster inference time.
 
The dataset was restructured to conform to the format requirements of Code Llama training~\cite{rozière2024codellamaopenfoundation}:
  \nopagebreak
  \begin{quote}
{"bug\_id": "ID",
"prompt":
"[INST] Generate detailed feedback in the format KM (Knowledge about Mistakes) and KH (Knowledge about how to proceed) for this Java code:
   \promptinsert{JAVA CODE}
[/INST]",  \\
"response": "KM: \promptinsert{KM response} `."KH: \promptinsert{KH response}`"}
\end{quote}
 The data was converted into JSON objects. To avoid overfitting and provide a robust picture of generalizability, we used 5-fold cross-validation, with each fine-tuned model instance trained on a subset of bug categories and then independently evaluated on the others.    
 %CodeLlamaTokenizerFast, a byte-level BPE tokenizer converts the data into tokens that are used for training and inference. The vocabulary size of the tokenizer is 32016 and the model weights are stored using 32-bit floating point numbers.
\begin{comment}
For instance, the tokenizer converts the following java statement into tokens as follows. tokenizer("System.out.println('Hello world')") 
{'input\_ids': [1, 2184, 29889, 449, 29889, 5248, 877, 10994, 3186, 1495], 'attention\_mask': [1, 1, 1, 1, 1, 1, 1, 1, 1, 1]}. The input\_ids represent the various tokens associated with the sequence, and attention\_mask is a binary vector that guides the model on which token to focus and which one to ignore. 
\end{comment}
Each token is converted into a vector of size 4096.
We employed QLoRA-style 4-bit quantization using nf4 format and double quantization on CodeLlama for this task-specific adaptation. During training the original model weights are frozen and the newly added weights are updated. This significantly reduces the number of training parameters from 6,743,789,568  to 5,242,880. The adapters were configured with a rank of 10, scaling factor 16 and a dropout rate of 0.08\%. LoRA adapters were applied to the q\_proj and v\_proj projection layers. This fine-tuning strategy adapts the base model for downstream tasks with lower computational resource requirements.
The prompt format is given in Prompt~\ref{prompt:Evaluation prompt}.
\begin{llmprompt}[label={prompt:Evaluation prompt}]
\small
{{PEFT Llama and baseline prompt  }}
[INST]Generate detailed feedback in the format  Knowledge about Mistakes (KM) and Knowledge about how to proceed (KH) for this Java code:
\promptinsert{JAVA CODE}[/INST]
\end{llmprompt}
\subsection{Prompt Engineering}
\label{sec:prompt-engineering}
We investigated the effectiveness of two prompting strategies, a baseline zero-shot prompt and in-context prompting, for feedback generation using CodeLlama-7b-Instruct-hf.
\begin{comment}
The baseline prompt format is given in Prompt~\ref{prompt:basic-llama}.
\begin{llmprompt}[label={prompt:basic-llama}]{{Baseline prompt}}
You are an expert Java programming tutor. Your goal is to provide detailed feedback to the student and make them aware of the mistakes in their code (Knowledge about mistakes) and how to fix them (Knowledge about how to proceed). \\

Provide feedback using the following format:  \\
         KM (Knowledge about mistakes): \\
         KH (Knowledge about how to proceed): \\
         
\promptinsert{JAVA CODE}
\end{llmprompt}
\end{comment}
For in-context prompting, we augmented the baseline prompt with three examples selected from the training dataset; see Prompt~\ref{prompt:in-contet-llama}.
\begin{llmprompt}[label={prompt:in-contet-llama}]{{In-context Llama prompt}}
\small
[INST]Generate detailed feedback in the format Knowledge about Mistakes (KM) and Knowledge about how to proceed (KH) for this Java code:
\promptinsert{JAVA CODE}.  Do NOT create additional examples.\\       
 Example 1    \\
   
    \promptinsert{JAVA CODE} \promptinsert{Feedback : KM, KH} \\
 Example 2     \\
\promptinsert{JAVA CODE} \promptinsert{Feedback : KM, KH} \\
  Example 3   \\
\promptinsert{JAVA CODE}  \promptinsert{Feedback : KM, KH}
[/INST]
\end{llmprompt}
A temperature value of 0.7, a maximum token limit of 8192, and a top-p (nucleus sampling) value of 0.7 were used. The feedback responses for the test dataset were collected using both prompting strategies.
\subsection{Evaluation}
\label{sec:evaluation-metrics}
%{\color{blue}
%We applied rubric-based manual annotations conducted by the first author, an instructor of CS1 courses, alongside automated metric analysis using BLEU~\cite{10.3115/1073083.1073135}, ROUGE~\cite{lin-2004-rouge}, and BERTScore~\cite{zhang2020bertscoreevaluatingtextgeneration}}. 
Our evaluation process has three components. The first is done from the instructor perspective, and involves rubric-based manual annotations conducted by the first author, an instructor of CS1 courses. To provide a more objective perspective, this is supported by the second component, which involves automated metric analysis. The third component, to evaluate the student perspective, is a small focus group discussion involving CS1 students.

For the instructor evaluation, we developed a rubric consisting of four binary criteria (see Table~\ref{tab:feedback-metrics}) where a score of 1 for a component indicates that the feedback perfectly aligns with  the specified characteristic. In cases with multiple KM/KH responses, each response is examined for misleading information and evaluated accordingly. Cross-annotation was performed by a second researcher to ensure consistency and reliability of the labelling of the KHH, KMA, MS and PA categories. This process prompted meaningful discussions and refined the labelling criteria.

For the objective evaluation, the automated evaluation metrics compute the similarity between the feedback produced by the models and the ground truth. BLEU~\cite{10.3115/1073083.1073135} evaluates how many exact word/n-grams in the output match those in the reference text. ROUGE~\cite{lin-2004-rouge}, on the other hand, measures how many n-grams from the reference text  are present in the generated output. BLEU emphasizes precision, whereas ROUGE focuses on recall. Among the ROUGE variants, we report the ROUGE-L as it captures the longest common subsequence between the generated and reference feedback. These two metrics rely on exact word/phrase overlap, and they may miss feedback that is semantically similar but lexically different. Therefore, BERTScore~\cite{zhang2020bertscoreevaluatingtextgeneration}, which captures the semantic similarity between the model response and the ground truth, was also employed. Specficially, we used BERTScore-F1, which balances precision and recall. Since feedback on the same idea can be expressed in multiple ways, evaluating it using multiple metrics enhances the robustness of the evaluation. The evaluation metrics were computed for all the samples in  the test dataset. 

For the student evaluation, we employed a mixed-methods evaluation following a small focus group discussion to investigate the student’s perception of different types of feedback for a CS1 Java programming assignments. The study was conducted following the institution's ethical approval. All undergraduate students in years 1 and 2 were invited to participate in this study. Eleven students signed the consent form, and seven students (P1 to P7) attended the meeting. Participants were provided with four buggy CS1-level Java code snippets and three different types of feedback to help them identify and fix the error in the code. The first feedback type was the compiler error message (E), the second  was feedback generated by ChatGPT (C), and the third feedback was produced by the fine-tuned model (F) used in this study. Participants rated each type of feedback from 1 to 5 across three factors: usefulness, clarity and structure.  In the second phase of the focus group discussion, the students were asked to reflect on their rating and discussed the rationale for it. The session was audio-recorded. We conducted a thematic analysis \cite{Braun01012006} of the transcripts to examine the students' perception of the feedback provided.  
 
\begin{table}
\centering
\footnotesize
%\begin{tabularx}{\linewidth}{l@{~}lX}
\begin{tabular}{lll}
  \toprule
  \multicolumn{2}{l}{\textbf{Metric}} & \textbf{Description} \\
\midrule
KMA & KM Accuracy & Addresses the core issue in the code \\
KHH & KH Helpfulness & Provides meaningful hint to the learner \\
MS & Misleading Suggestions & Wrong fix or irrelevant advice \\
PA & Prompt Adherence & Follows the specified structure (KM, KH) \\
\bottomrule
\end{tabular}
\caption{Descriptions of feedback evaluation metrics.}
 \label{tab:feedback-metrics}
\end{table}

\section{Results, Discussion, and Limitations}
\label{sec:results-discussion}
The following section outlines the findings from our experiments involving PEFT and prompt engineering. We applied manual and automated evaluation metrics    along with student evaluations to assess the quality of the generated responses.

%Although Code Llama  excels at generating code, its effectiveness in producing pedagogically meaningful KM/KH responses has not been sufficiently explored. We evaluated the Code Llama baseline against the PEFT-enhanced model and also assessed the impact of few-shot prompt engineering on the baseline model. 
 
\subsection{Instructor Evaluation} 
\begin{figure}[htbp]
\centering
\begin{tikzpicture}
\begin{axis}[
    ybar,
    bar width=4pt,
    width=9 cm,  % fits one column nicely
    height=5.5cm,
    ymin=0,
    ymax=100,
    ylabel={Score (\%)},
    symbolic x coords={KM Accuracy, KH Helpfulness, Misleading Suggestions,Prompt Adherence},
    xtick=data,
    xtick style={draw=none},
    xticklabel style = {rotate=45, anchor=east},
    ymajorgrids=true,
    grid style=dashed,
    legend style={
        at={(0.5,1.2)},
        anchor=north,
        font=\scriptsize,
        legend columns=2
    },
    nodes near coords,
    every node near coord/.append style={font=\tiny},
    tick label style={font=\scriptsize},
    label style={font=\scriptsize},
]

% Baseline
\addplot[fill= lightblue] coordinates {
    (KM Accuracy,20)
    (KH Helpfulness,26) 
     (Misleading Suggestions,83)
     (Prompt Adherence,54)
};

% PE-FewShot
\addplot [fill = darkerblue] coordinates {
    (KM Accuracy,54)
    (KH Helpfulness,46)
    (Misleading Suggestions,60)
     (Prompt Adherence,86)
};
% PEFT
\addplot [fill=yellow!30] coordinates {
    (KM Accuracy,61)
    (KH Helpfulness,60)
      (Misleading Suggestions,47)
     (Prompt Adherence,95)
};
\legend{Baseline, PE-FewShot, PEFT}
\end{axis}
\end{tikzpicture}
\caption{Comparison of KM accuracy, KH helpfulness, number of misleading suggestions (lower is better), and prompt adherence across model configurations. Baseline refers to the standard Llama response. PE-FewShot refers to prompt engineering with few-shot examples. PEFT refers to the fine-tuned model.}
\label{fig:km-kh-grouped}
\end{figure}
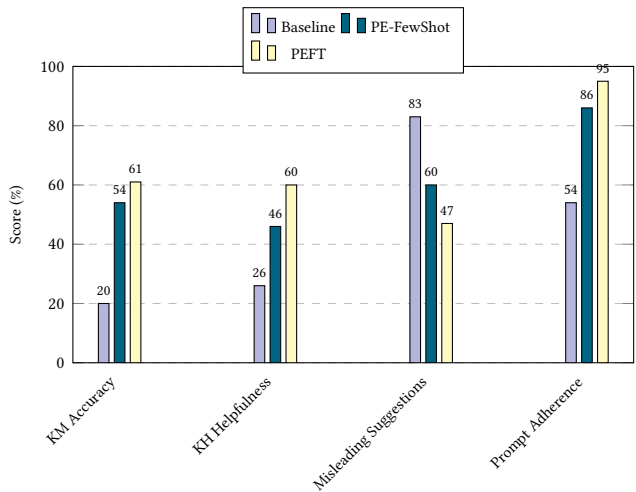

  Figure~\ref{fig:km-kh-grouped}  summarizes the rubric-based  comparative analysis conducted by the first author (CS1 course instructor). Our results show that the fine-tuned model (CL7B-LoRA-JavaFB) outperformed the other techniques on all of the four rubric criteria (see Table~\ref{tab:feedback-metrics}). Examples of feedback produced by each model, along with the dataset, are provided in the footnote \footnote{\url{https://anonymous.4open.science/r/JCODE_KM_KH-4BEC}}.
  %using the rubric. Figure~\ref{fig:km-kh-grouped} 
  %presents a comparative analysis using our rubric-based manual annotations of feedback quality

\begin{comment}
 \begin{figure}[htbp]
\centering
\begin{tikzpicture}
\begin{axis}[
    ybar,
    bar width=6pt,
    width=7.8cm,  % fits one column nicely
    height=5.5cm,
    ymin=0,
    ymax=100,
    ylabel={Score (\%)},
    symbolic x coords={Misleading Suggestions, Prompt Adherence},
    xtick=data,
    xtick style={draw=none},
    ymajorgrids=true,
    grid style=dashed,
    legend style={
        at={(0.5,-0.3)},
        anchor=north,
        font=\scriptsize,
        legend columns=2
    },
    nodes near coords,
    every node near coord/.append style={font=\tiny},
    tick label style={font=\scriptsize},
    label style={font=\scriptsize},
]

% Baseline
\addplot[fill= lightblue]  coordinates {
    (Misleading Suggestions,58)
     (Prompt Adherence,76)
};
% PE-Zeroshot
 \addplot [fill =pastel] coordinates {
    (Misleading Suggestions,52)
     (Prompt Adherence,95)
};
% PE-FewShot
\addplot [fill = darkerblue] coordinates {
    (Misleading Suggestions,26)
     (Prompt Adherence,95)
};
% PEFT
\addplot [fill=yellow!30] coordinates {
    (Misleading Suggestions,11)
     (Prompt Adherence,98)
};

\legend{Baseline, PE-Zeroshot, PE-FewShot, PEFT}
\end{axis}
\end{tikzpicture}
\caption{Comparison of Misleading Suggestions and Prompt Adherence Baseline, PE-Zeroshot, PE-FewShot, and PEFT configurations.}
\label{fig:mislead-prompt}
\end{figure}
\end{comment}
%Figure~\ref{fig:km-kh-grouped} presents a comparative analysis using our rubric-based manual annotations of feedback quality. 
The fine-tuned model demonstrated superior KM accuracy, achieving 61\% compared to 20\% for the un-tuned baseline model. In terms of KH helpfulness, it achieved a score of 60\% compared to 26\%. We also observed that the rate of  misleading suggestions dropped from 83\% to 47\%, and  prompt adherence increased from 54\% to 95\%. These results demonstrate the potential of PEFT in generating pedagogically meaningful feedback.
 
For the prompt engineering strategies, we see that in-context learning achieved  better performance than the baseline model in KM accuracy  scores with 54\% over 20\%. Moreover, KH helpfulness increased from 26\% to 46\% and the misleading suggestions dropped to 60\%. This shows that prompt engineering with task-specific examples can also improve  feedback quality. Both strategies consistently achieved high PA scores.

\begin{comment}
The following are the sample example feedback generated by each of the techniques. For the following code snippet, 
{\color{blue}Change!!!-Baseline vs LoRA model response comparison}
\begin{tcolorbox}[title={PEFT model response}]
\texttt{   int[] grades = new int[] {90, 85, 77}; \\ 
   List gradesList = grades; \\ }
The compiler error message is "Type mismatch: cannot convert from int[] to List". This might be challenging for a novice programmer to understand. \\ The fine-tuned model response is as follows 
KM : " Assigning an array to a List variable directly, which is not possible because arrays are not assignable to Lists."
and the corresponding KH response is " Declare a List and use the List's add method to add each element of the array to the List. Alternatively, use the Arrays.asList method to create a List from the array." \\
Groundtruth - KH is "Convert the array to a List using Arrays.asList() and ensure the List is parameterized with the correct type (e.g., List<Integer>)".  
\end{tcolorbox} 
\end{comment}

%\subsubsection*{Example}

Overall, PEFT demonstrated the strongest performance and produced the highest-quality feedback compared to other approaches. Additionally, the evaluation revealed that the PEFT model introduced minimal redundancy and produced the least number of misleading statements. This is an essential consideration in a pedagogical environment, as these elements directly affect student performance. 
The evaluation also assessed the performance of the PEFT model within the two error categories, imperative and object-oriented. It achieved an accuracy of 49\% for imperative errors and 52\% for object-oriented errors, suggesting that performance across error categories is relatively consistent.
%, in comparison to baseline model, which achieved an accuracy of less than 20\%.  

\begin{comment}
The dataset comprises multiple bug types and feedback categories, however only a subset of these categories are present in the  test set. To ensure the robustness of our findings in the context of diverse bug types and feedback categories, each configuration was trained and evaluated across multiple random seeds (42, 41, and 7). Performance outcomes from the multiple seed runs were aggregated to compute average performance.  KM scores range from 56\% to 67\%, with a mean of 62\%, demonstrating that the model reliably detects coding mistakes across diverse bug categories. The KH metric ranges from 60\% to 68\%, with a mean of 63\%, reflecting reliability in generating quality feedback.  MS scores is between 36\% and 52\%, with a mean of 45\% showing that the some variation in the results. PA scores are high across all runs, indicating that the model respects the overall structure. Among the three categories the model performs best on Object-oriented concept related errors with an accuracy of 68\%, followed by misconceptions at 52\%, and imperative concepts at 49\%. 
\end{comment}%The process involved cross-checking the responses against a ground truth to ensure accuracy.

\subsection{Objective Evaluation} 
\begin{figure}[th]
\centering

% ================= KM plot =================
\begin{minipage}{0.48\linewidth}
\centering
\begin{tikzpicture}
\begin{axis}[
    ybar,
    bar width=4pt,
    width=\linewidth,
    height=5.5cm,
    ymin=0,
    ymax=1,
    ylabel={Score},
    symbolic x coords={KM-BLEU, KM-ROUGE, KM-BERTScore},
    xtick=data,
    xtick style={draw=none},
    xticklabel style={rotate=45, anchor=east},
    ymajorgrids=true,
    grid style=dashed
]

% Baseline
\addplot[fill=lightblue] coordinates {
    (KM-BLEU,0.023) (KM-ROUGE,0.22) (KM-BERTScore,0.87)
};

% PE-FewShot
\addplot[fill=darkerblue] coordinates {
    (KM-BLEU,0.02) (KM-ROUGE,0.22) (KM-BERTScore,0.87)
};

% PEFT
\addplot[fill=yellow!30] coordinates {
    (KM-BLEU,0.04716) (KM-ROUGE,0.2924) (KM-BERTScore,0.89192)
};

\end{axis}
\end{tikzpicture}
\subcaption{KM Feedback}
\end{minipage}
\hfill
% ================= KH plot =================
\begin{minipage}{0.48\linewidth}
\centering
\begin{tikzpicture}
\begin{axis}[
    ybar,
    bar width=4pt,
    width=\linewidth,
    height=5.5cm,
    ymin=0,
    ymax=1,
    ylabel={Score},
    symbolic x coords={KH-BLEU, KH-ROUGE, KH-BERTScore},
    xtick=data,
    xtick style={draw=none},
    xticklabel style={rotate=45, anchor=east},
    ymajorgrids=true,
    grid style=dashed
]

% Baseline
\addplot[fill=lightblue] coordinates {
    (KH-BLEU,0.03) (KH-ROUGE,0.19) (KH-BERTScore,0.86)
};

% PE-FewShot
\addplot[fill=darkerblue] coordinates {
    (KH-BLEU,0.03) (KH-ROUGE,0.23) (KH-BERTScore,0.87)
};

% PEFT
\addplot[fill=yellow!30] coordinates {
    (KH-BLEU,0.05632) (KH-ROUGE,0.30934) (KH-BERTScore,0.8893)
};

\end{axis}
\end{tikzpicture}
\subcaption{KH Feedback}
\end{minipage}

% -------- Shared legend --------
\vspace{0.4em}
\begin{tikzpicture}
\begin{axis}[
    hide axis,
    xmin=0, xmax=1,
    ymin=0, ymax=1,
    legend columns=3,
    legend style={
        at={(0.5,0.5)},
        anchor=center
    }
]
\addlegendimage{area legend,fill=lightblue}
\addlegendentry{Baseline}

\addlegendimage{area legend,fill=darkerblue}
\addlegendentry{PE-FewShot}

\addlegendimage{area legend,fill=yellow!30}
\addlegendentry{PEFT}
\end{axis}
\end{tikzpicture}
\captionsetup[figure]{font=small}
\caption{Comparison of BLEU, ROUGE, and BERTScore metrics for KM and KH feedback across different model configurations.}
\label{fig:comparison-merged}
\end{figure}
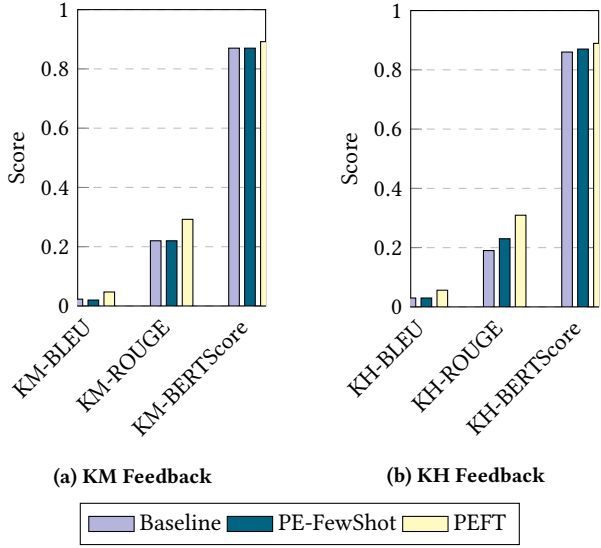

To complement the human annotated evaluations, we conducted an automated analysis of feedback using BLEU, ROUGE, and BERTScore. Figure~\ref{fig:comparison-merged} summarize the results.  Generally, we see the same pattern as in the manual analysis, with the fine-tuned model performing best on these metrics, followed by in-context learning, and then the baseline model. Although the models demonstrated low BLEU and ROUGE scores, the high BERTScore indicates that they generated meaningful feedback semantically similar to the ground truth. 
The fine-tuned model achieved particularly high BERTScore values indicating strong semantic alignment. 
While these metrics provide valuable insights, they do not capture aspects such as clarity, format adherence or the pedagogical soundness of the feedback. Our observations also indicate that high BERTScores do not guarantee pedagogically soundness, cautioning against the use of automated metrics alone to evaluate model performance. For instance, while the baseline model achieved a BERTScore of 0.84, manual checking of the responses reported an accuracy rate of only 20-26\%.

\begin{comment}
 \begin{figure} [th] 
\centering
\begin{tikzpicture}
\begin{axis}[
    ybar,
    bar width=7pt,
    width=15cm,
    height=9cm,
    enlarge x limits=0.15,
    ymin=0,
    ymax=1,
    ylabel={Score},
    symbolic x coords={
        KM-BLEU, KM-ROUGE, KM-BERTScore,
        KH-BLEU, KH-ROUGE, KH-BERTScore
    },
    xtick=data,
    xtick style={draw=none},
    x tick label style={rotate=45, anchor=east},
    ymajorgrids=true,
    grid style=dashed,
    legend style={at={(0.5,-0.25)}, anchor=north, legend columns=4}     
]

% Baseline
\addplot+[fill=gray!40!white] coordinates {
    (KM-BLEU,.02) (KM-ROUGE,.16) (KM-BERTScore,.84)
    (KH-BLEU,.01) (KH-ROUGE,.12) (KH-BERTScore,.83)
};

% PE-Zeroshot
\addplot+[fill=violet!10!white] coordinates {
    (KM-BLEU,0.04) (KM-ROUGE,.27) (KM-BERTScore,.88)
    (KH-BLEU,.03) (KH-ROUGE,.23) (KH-BERTScore,.87)
};

% PE-FewShot
\addplot+[fill=cyan!30!white] coordinates {
    (KM-BLEU,0.04) (KM-ROUGE,0.33) (KM-BERTScore,0.89)
    (KH-BLEU,0.037) (KH-ROUGE,0.26) (KH-BERTScore,0.88)
};
% PEFT
\addplot+[fill=blue!20!white] coordinates {
    (KM-BLEU,0.09) (KM-ROUGE,0.36) (KM-BERTScore,0.90)
    (KH-BLEU,.07) (KH-ROUGE,.35) (KH-BERTScore,.89)
};
\legend{
  Baseline,
  PE-Zeroshot,
  PE-FewShot,
  PEFT
}
 \end{axis}
\end{tikzpicture}
\caption{Comparison of BLEU, ROUGE, and BERTScore metrics for KM and KH feedback across different model configurations.}
\label{fig:comparison-metrics}
\end{figure}
\end{comment}
 
Although the rate of misleading suggestions generated by the PEFT-model is  significantly lower than the other techniques, this remains a concern and underscores the necessity of verifying results before deploying it in an educational setting. 
We  observed that the generated feedback tend to be prescriptive in nature, offering direct instructions. 
This feedback can be enriched by integrating reasoning elements, scaffolding questions or partial hints to enhance its pedagogical effectiveness. Currently the PEFT model is trained with <Code, KM, KH> triplets. Incorporating task description into the training data is expected to improve the performance.

\subsection{Student Evaluation}
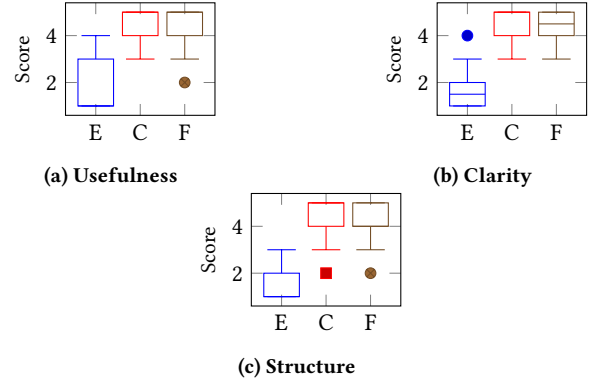
\begin{figure}
\centering
\begin{subfigure}[b]{0.20\textwidth}
\centering
\begin{tikzpicture}
\begin{axis}[
    boxplot/draw direction=y,
    ylabel={\small Score},
    xtick={1,2,3},
    xticklabels={E,C,F},
    width=\textwidth
]
\addplot+[boxplot] table[row sep=\\, y index=0] {3\\3\\1\\1\\3\\2\\1\\1\\2\\1\\1\\1\\3\\3\\1\\1\\1\\3\\1\\1\\4\\4\\1\\1\\1\\1\\1\\};
\addplot+[boxplot] table[row sep=\\, y index=0] {4\\4\\4\\4\\5\\4\\5\\5\\4\\5\\5\\5\\4\\4\\3\\3\\5\\5\\5\\5\\5\\5\\4\\5\\5\\5\\5\\5\\};
\addplot+[boxplot] table[row sep=\\, y index=0] {3\\4\\4\\4\\2\\5\\4\\4\\5\\5\\5\\5\\4\\5\\4\\5\\4\\5\\5\\5\\5\\5\\5\\5\\4\\4\\4\\5\\};
\end{axis}
\end{tikzpicture}
\caption{Usefulness}
\label{fig:usefulness}
\end{subfigure}
\hfill
\begin{subfigure}[b]{0.20\textwidth}
\centering
\begin{tikzpicture}
\begin{axis}[
    boxplot/draw direction=y,
    ylabel={\small Score},
    xtick={1,2,3},
    xticklabels={E,C,F},
    width=\textwidth
]
\addplot+[boxplot] table[row sep=\\, y index=0] {3\\2\\1\\1\\2\\4\\2\\1\\3\\2\\1\\1\\2\\2\\1\\1\\2\\2\\1\\1\\3\\3\\1\\1\\1\\2\\1\\1\\};
\addplot+[boxplot] table[row sep=\\, y index=0] {5\\5\\5\\5\\4\\3\\5\\5\\4\\5\\5\\5\\4\\3\\4\\3\\5\\5\\4\\5\\5\\5\\5\\5\\5\\5\\5\\5\\};
\addplot+[boxplot] table[row sep=\\, y index=0] {3\\4\\5\\5\\3\\4\\4\\5\\4\\5\\5\\5\\3\\4\\4\\4\\5\\5\\5\\5\\5\\3\\5\\5\\4\\4\\3\\5\\};
\end{axis}
\end{tikzpicture}
\caption{Clarity}
\label{fig:clarity}
\end{subfigure}
\hfill
%\vspace{1em} % spacing before next row
\begin{subfigure}[b]{0.20\textwidth}
\centering
\begin{tikzpicture}
\begin{axis}[
    boxplot/draw direction=y,
    ylabel={\small Score},
    xtick={1,2,3},
    xticklabels={E,C,F},
    width=\textwidth
]
\addplot+[boxplot] table[row sep=\\, y index=0] {2\\2\\1\\1\\1\\3\\1\\1\\2\\1\\1\\1\\2\\2\\1\\1\\1\\2\\1\\1\\3\\3\\1\\1\\1\\1\\1\\1\\};
\addplot+[boxplot] table[row sep=\\, y index=0] {5\\5\\5\\5\\4\\5\\5\\5\\4\\4\\5\\4\\3\\3\\3\\2\\5\\5\\3\\5\\4\\5\\4\\4\\5\\5\\5\\5\\};
\addplot+[boxplot] table[row sep=\\, y index=0] {3\\4\\4\\4\\2\\5\\3\\4\\5\\5\\5\\5\\4\\4\\4\\4\\4\\4\\5\\5\\4\\4\\5\\5\\4\\4\\4\\5\\};
\end{axis}
\end{tikzpicture}
\caption{Structure}
\label{fig:structure}
\end{subfigure}
\caption{Distributions of student-assessed scores for different feedback types: compiler error message (E), ChatGPT (C), and fine-tuned model (F).}
\label{fig:feedback_boxplots}
\end{figure}

Based on  the positive evaluation of the PEFT model, we carried out a student evaluation, where participants assessed the PEFT model responses alongside those from a proprietary model. The findings indicate that  the feedback  generated by ChatGPT (C) and the model (F) received positive responses across all three parameters, as shown in Figure~\ref{fig:feedback_boxplots}.
%% s~\ref {fig:usefulness}, \ref {fig:clarity} and \ref {fig:structure}. 
Participants unanimously agreed that compiler error messages (E) were ineffective. %An example of feedback samples is shown in the footnote \footnote{\url{https://anonymous.4open.science/r/JCODE_KM_KH-4BEC}}.
The thematic analysis of the transcripts identified five key themes: clarity of explanation,
level of detail, usefulness, 
actionability of feedback, and  the
presence or absence of code.   

\subsubsection{Theme - Clarity of explanation,
Level of detail}
Most participants noted that, while Feedback C was helpful, it was often overly verbose, which may be overwhelming for beginners. E.g., \textit{"detailed feedback but too many options"} (P1), \textit{"too much information"} (P4), \textit{"bit too wordy"} (P5), \textit{"overexplain the concepts"} (P6).
However, P4 stated,  \textit{"I didn’t feel like it was a lot of information. When I’m learning, I feel that I want to know why there’s a problem, and what other options and additional information are available."}. Results indicate that learners’ preferred level of feedback detail may affect their learning experience.  Another observation was that \textit{"Feeback F would be more appropriate for learners with basic knowledge of Java"} (P1), \textit{"and is best suited for   formative assessments in a lab setting, whereas Feedback C might be more appropriate for summative assessment feedback"} (P5).
\subsubsection{Theme - Usefulness, Actionability of feedback}
Aligned with these observed preferences, all participants agreed that actionable feedback promotes engagement and motivates learners to tackle complex challenges. P7 stated, \textit{"Feedback C clearly identified the error and explained why the error was there, and how to fix it."} and \textit{"Feedback F was precise, straight to the point explanation, encouraging student to think of solution, and directs them to the correct path."} There was also a suggestion to update Feedback F with additional explanations for technical terms. P4 commented, \textit{"Some phrases such as `incompatible' types could be explained more. I felt I could work with it and get the code fixed."}
\subsubsection{Theme - Presence or absence of code}
Participants observed that Feedback C often provides the correct solution, which could lead to over-reliance and limited independent learning. For instance, P3 noted that, although \textit{"Feedback C was good, but the correct solution might result in some students skipping the feedback and copying the code"}.  Participants agreed that early access to the correct code can negatively affect critical thinking.

Overall, the findings indicate that participants valued the feedback generated by the fine-tuned model, and its ratings were comparable to those of the proprietary model. According to the participants, Feedback F is well-suited for a lab setting, as it encourages students to engage in critical thinking and problem-solving. Our findings are consistent with those reported in \cite{solano2025narrowinggapsupervisedfinetuning}. 

 \subsection{Limitations} \label{sec:limitations}
Our dataset was generated using Deepseek-R1; therefore, the model performance is influenced and limited by the original model. Additionally, the dataset used in this study is limited in size, which may affect the generalizability of the results. Furthermore, this analysis is confined to Java and a specific set of bugs identified in the literature. The student evaluation was based on a small focus group; additional research with larger samples would be useful to verify that the findings generalize.   %The rubric-based evaluation was conducted by the first author, which may influence the results. The student evaluation was based on a small focus group; additional research with larger samples would be useful to verify that the findings generalize.}

\section{Conclusions}
\label{sec:conclusions}
LLMs offer potential in providing guidance to student learners, but the financial constraints of proprietary models and the computational demands of large-parameter architectures pose significant challenges in educational contexts. Open models, which are significantly smaller and freely deployable, present a viable alternative, but are generally less capable than proprietary models. 
%For instance, we found that they frequently provided full code fixes, which could hinder the learning process, alongside generic responses and extra information that could confuse students. 
To address this, in our work we have shown how the capabilities of a more capable model can be distilled into an open model by the use of synthetic data generation and parameter-efficient fine-tuning. 
%The baseline model produce generic responses that includes comments about coding style and Java naming conventions along with the actual issue in the code. Moreover, providing  the full functional version of the code along with the feedback hinders the learning process. \\
%Proprietary models restrict customization options and, therefore, present significant challenge in providing pedagogically aligned feedback.
These fine-tuned models offer significantly better guidance to students, in terms of providing actionable feedback on buggy Java programs, than the baseline model. Notably, the accuracy and helpfulness of the feedback approximately doubles after fine-tuning, and there is also considerable improvement in terms of reducing hallucinations. We also show that fine-tuning is more effective in this domain than prompt engineering, although the latter approach also offers improvement over the baseline model. The presence of misleading information in PEFT-based feedback cannot be ignored, highlighting the need for human evaluation in education settings. As part of future work, we plan to extend the dataset by including more diverse feedback covering various Java programming concepts and to optimise the model performance and effectiveness.
\bibliographystyle{ACM-Reference-Format}
\bibliography{reference}
 
\end{document}